\documentclass[10pt,conference,a4paper]{IEEEtran}
\usepackage{cite}
\usepackage{amsmath,amssymb,amsfonts}
\usepackage{algorithmic}
\usepackage{graphicx}
\usepackage{textcomp}
\usepackage{xcolor}
\usepackage[utf8]{inputenc}
\usepackage{caption}
\usepackage[hang,scriptsize,tight,nooneline]{subfigure}
\usepackage{subcaption}
\usepackage{amsmath,amsfonts}
\usepackage{algorithmic}
\usepackage{graphicx}
\usepackage{multirow}
\usepackage{textcomp}
\usepackage{xcolor}
\usepackage{float}
\usepackage{xspace}
\usepackage{multirow}
\usepackage{enumitem}
\usepackage{csquotes}

\usepackage[font=small,labelfont=bf,textfont=it]{caption}

\newcommand{\simulator}{\textsf{\small{\mbox{\textit{xeoverse{}~}}}}}
\newcommand{\simulatorPython}{\textsf{\small{\mbox{xeoverse Back Stage}~}}}
\newcommand{\simulatorMininet}{\textsf{\small{\mbox{xeoverse Main Stage}~}}}

\begin{document}

\title{$x$eoverse: A Real-time Simulation Platform for Large LEO Satellite Mega-Constellations~\vspace{-0.12in}}
% ~\vspace{-0.09in}
% \author{\IEEEauthorblockN{Mohamed M. Kassem}
% \IEEEauthorblockA{\textit{University of Surrey} \\
% % City, Country \\
% % email address or ORCID
% }
% \and
% \IEEEauthorblockN{Nishanth Sastry}
% \IEEEauthorblockA{\textit{University of Surrey} \\
% % City, Country \\
% % email address or ORCID
% }

% }

\author{\IEEEauthorblockN{Mohamed M. Kassem,
Nishanth Sastry}
\IEEEauthorblockA{University of Surrey\\
\{m.kassem, n.sastry\}@surrey.ac.uk~\vspace{-0.2in}}}

\IEEEoverridecommandlockouts

\maketitle

\begin{abstract}

In the evolving landscape of satellite communications, the deployment of Low-Earth Orbit (LEO) satellite constellations promises to revolutionize global Internet access by providing low-latency, high-bandwidth connectivity to underserved regions. However, the dynamic nature of LEO satellite networks, characterized by rapid orbital movement and frequent changes in Inter-Satellite Links (ISLs), challenges the suitability of existing Internet protocols designed for static terrestrial infrastructures. Testing and developing new solutions and protocols on actual satellite mega-constellations are either too expensive or impractical because some of these constellations are not fully deployed yet. This creates the need for a realistic simulation platform that can accurately simulate this large scale of satellites, and allow end-to-end control over all aspects of LEO constellations. 

This paper introduces \simulator, a scalable and realistic network simulator designed to support comprehensive LEO satellite network research and experimentation. By modeling user terminals, satellites, and ground stations as lightweight Linux virtual machines within Mininet and implementing three key strategies -- pre-computing topology and routing changes, updating only changing ISL links, and focusing on ISL links relevant to the simulation scenario -- \simulator achieves real-time simulation, where 1 simulated second equals 1 wall-clock second. Our evaluations show that \simulator outperforms state-of-the-art simulators Hypatia and StarryNet in terms of total simulation time by being 2.9 and 40 times faster, respectively.

\end{abstract}

\vspace{-0.06in}
\section{Introduction}
\vspace{-0.01in}

The rapid advancement in satellite and space technology, especially with the development of reusable rockets and the extensive deployment of Low-Earth Orbit (LEO) satellites, is transforming the way we access the Internet from space. Unlike the traditional geostationary (GEO) satellites stationed  $35,000$ KM above Earth, LEO satellites orbit at much lower altitudes, between $500$ and $2,000$ KM~\cite{zee2013theory, geosatPerdices}, reducing end-to-end latency significantly. This new \textit{space race}, led by Starlink, OneWeb, and Kuiper, aims to provide low-latency, high-bandwidth Internet coverage across the globe, including the 85\% of Earth's surface that currently lacks reliable connectivity (a drawback which has impeded adoption of traditional broadband applications even in developed countries~\cite{karamshuk2015factors}). 

However, the dynamic nature of LEO satellite networks, with their high-speed movement and frequent changes in Inter-Satellite Links (ISLs), poses a challenge to the existing terrestrial Internet protocols and algorithms, which were designed for a static infrastructure. For instance, measurements have shown the traditional congestion control protocols do not perform optimally on Starlink~\cite{kassem2022browser}. 

This creates a dire need for simulation platforms that allow end-to-end control over all aspects of LEO megaconstellations-based architectures. It is challenging to create a \textbf{high-fidelity} platform that \textit{(i)} \textit{faithfully replicates details} such as satellite movements at 27,000 KM/hr, the associated frequent link changes, the detailed link representation in terms of RF parameters and signal-to-noise ratio (SNR) and the effect of local weather conditions on links across the globe, \textit{(ii)} can do this whilst \textbf{scaling} to the thousands of satellites which are already deployed on constellations such as Starlink and the tens of thousands of possible links, while at the same time being \textit{(iii)} \textbf{responsive} -- enabling fast simulations that do not slow down with the growing scale of the constellations and the fidelity of the simulation. 
Indeed, we ideally want a simulator that operates in \textit{real-time}, i.e., one second of simulation time is executed in one second of wall-clock time, without compromising on the fidelity or the scalability. To be accessible, we also want \textit{(iv)} a \textbf{low footprint} simulator, that does not require too many machines and high compute power. 

Our main contribution in this paper is the design and implementation of \simulator, a high-fidelity real-time simulator that can simulate the entire current Starlink constellation (5,442 satellites) on a single 26-cores 64GB machine. Although we focus on Starlink because it is currently the largest constellation available, our simulator can be adapted to any other constellation where the positions of the satellites can be described using the standard Two-Line Element (TLE) Set \cite{tlefiles}.

\simulator relies on three key observations to create a scalable yet responsive architecture without compromising on simulation fidelity: The first observation is that while the movement of satellites causes significant changes in the global topology of LEO constellations across different time points, these movements are \textit{predictable} since the satellites are in fixed orbits. Thus, we can \textit{pre-compute} the topology for each time point of interest in the simulation, ahead of the running the actual simulation itself. Then, as long as the topology for a given time point can be instantiated in real-time, we can ensure the whole simulation proceeds in real-time.  

Our second and third observations help to efficiently manage the topology changes from one time point to another: The second observation is that although there is constant link churn due to satellite movement, most links stay the same over small time periods. Thus link changes can be a manageably small in number from one time point to the very next time point in high fidelity simulations that require high time resolution (we support time changes down to 1ms, although FCC filings suggest that Starlink satellite handovers happen on the order of 15 seconds or more\cite{starlink15s}). \simulator scales by computing  the subset of links which change from one time point to the next. The third observation goes further by noticing that in many simulation use cases, the users are only interested in a subset of the entire constellation (e.g., a train operator using the simulator to understand whether Starlink or OneWeb offers better coverage over a train route in France may not need all details of satellites over North America to be updated). Given a traffic matrix of flows of interest, \simulator computes which elements (user terminals, satellites, and gateways) of the entire constellation carry this traffic of interest, and applies dynamic topology updates only for those elements, greatly enhancing scalability and decreasing compute footprint.

We compare \simulator with two state-of-the-art alternatives: Hypatia~\cite{kassing2020exploring} and StarryNet~\cite{lai2023starrynet}. We focus on a low footprint setting, using only one machine. Our evaluations show that when simulating one flow over the entirety of Starlink Shell 1, \simulator is more responsive than both: it is $40$x times faster than StarryNet, and $2.3$x times faster compared to Hypatia. \simulator scalability advantages become more apparent as the number of flows are increased -- In our single machine setting, StarryNet fails to scale beyond around 1023 satellites, thus failing to simulate even Shell 1 of Starlink. Hypatia slows down, taking 5x times more time than \simulator in the simulation updates phase. Furthermore, \simulator adds more link-level details than both Hypatia and StarryNet, allowing us to show differences between different weather conditions, similar to that observed on a real node. \simulator predicted throughput is within 16\% (2\%) of observed throughput during rainy (clear weather) conditions. Over a long distance link (London--San Francisco), \simulator is able to better reflect the changes in throughput due to temporary disruptions of Inter Satellite Links as compared to Hypatia (StarryNet is unable to scale to this scenario due to the number of satellites involved).

\vspace{-0.06in}
\section{Related Work}
\label{sec:relatedwork}

Network simulators such as \simulator serve as essential tools for examining network performance, testing new protocols, and applications on a large scale within a controlled, realistic environment before real-world deployment. NS-3, and its predecessor NS-2, are examples of such tools, which have enabled a large amount of networking research. NS-3\cite{henderson2008network}, a discrete-event network simulator has numerous modules (e.g., WiFi, 4G, 5G) developed atop it to expand its utility. Mininet\cite{de2014using}, designed with a focus on Software-Defined Networking (SDN), represents another simulation platform. However, both NS-3 and Mininet are general-purpose network simulators and are not specifically tailored for simulating LEO mega-constellations of satellites.

Over recent years, the interest in LEO satellite networks has surged, with efforts broadly categorized into three areas based on realism: real-world constellation deployment, measurement campaigns, and simulation platforms. 

\textbf{Real-world constellation development.} The Tiansuan constellation~\cite{wang2021tiansuan} adopts the stance that researchers need their own real-world constellation deployment for experimentation and full control over all aspects of the network. Despite the high-fidelity of this approach, it also comes with significantly high cost. To the best of our knowledge, only two Tiansuan satellites have been launched so far (http://www.tiansuan.org.cn/morenews.html, accessed 10 May 2024), and the plan is to launch up to 6 satellites in the first phase. It remains hard to match this to the scale of commercial closed mega-constellations e.g., Starlink, which has thousands of satellites. Thus, various important questions such as ISLs are difficult to study in this system. 

\textbf{Real-world measurement campaigns.} Unfortunately, researchers do not have direct access to the satellites of commercially deployed mega constellations such as Starlink. Therefore, given the barriers to setting up a parallel real-world research constellation, the research community has made notable strides in real-world measurement studies on live and operational commercial LEO satellite networks~\cite{ma2023network, raman2023dissecting, kassem2022browser, michel2022first, mohan2023multifaceted, izhikevich2023democratizing}. Such studies focus mainly on Starlink, with other constellations (e.g., OneWeb, Telesat) being less explored. These efforts, in addition, are limited to data from end devices (i.e., user terminals), and lack direct insights into the network's inner workings, such as gateways or satellites. This limitation highlights the persistent need for simulation software. While these measurement efforts treat the constellation as a black box, simulations allow in-depth exploration of design decisions within the constellation itself, such as satellite inter-connectivity and queueing disciplines, etc. These simulations are crucial for deeply understanding the complexities of LEO satellite mega-constellations.

\textbf{LEO Simulation Tools.} Most of the early work on LEO networks such as \cite{bhattacherjee2019network, handley2018delay} developed custom-built simulators to address specific research questions. More recently, general-purpose LEO simulators such as Hypatia\cite{kassing2020exploring}, Celestial\cite{pfandzelter2022celestial}, and StarryNet\cite{lai2023starrynet} have been developed. Hypatia, built on top of NS-3, provides packet-level analysis but faces scalability challenges due to computational overhead. Celestial, developed using SILLEO-SCNS, is exclusively focused on edge computing within LEO networks. StarryNet utilizes Docker virtualization to simulate satellite and ground networks; however, it encounters scalability limits due to Docker's constraints. StarryNet and Hypatia align most closely with our approach, aiming to simulate entire mega-constellations comprehensively. We discuss these approaches in detail in \S\ref{sec:design} and evaluate our simulator against Hypatia and StarryNet in \S\ref{sec:evaluation}.

\textbf{Other Simulation Tools.} High-fidelity link simulators, such as MATLAB SIMULINK~\cite{matlab_sim} and ATK STK~\cite{stk_agi}, provide detailed simulations of ground-to-satellite communication links. These tools consider various factors affecting communication quality, including signal loss, the Doppler effect, modulation techniques, antenna types, beams and radio frequency (RF) parameters. While effective for modeling individual links, these simulators are not designed to manage the complexity of entire mega-constellations. They focus on the simulation of communication links rather than the overall network performance or the dynamic interactions within large LEO satellite networks. Consequently, while useful for detailed link analysis, these tools do not offer insights into the broader network characteristics and performance of satellite constellations.

\section{Xeoverse's Design Approach}
\label{sec:design}

\subsection{Desiderata}
\label{sec:desiderata}
\vspace{-.05in}
In developing a simulator for LEO mega-constellation networks, it is imperative to outline key design requirements that ensure its effectiveness. These are foundational to replicating the complexity and dynamics of satellite networks. We start first by defining these desiderata, and then discuss how we achieved these characteristics in \simulator:

\begin{enumerate}[itemsep=0pt, parsep=0pt, topsep=0pt]
\renewcommand{\labelenumi}{\textbf{D\arabic{enumi}.}}

    \item \textbf{Scalability:} The need for extreme scalability distinguishes LEO networks from Geostationary satellite networks, as LEO requires thousands of satellites for global coverage compared to GEO -- Starlink \textit{Shell 1} has $1584$ satellites, and there are $5400+$ satellites overall. This mega constellation size presents a significant computational challenge for simulations, as accurately representing the complex dynamics of LEO networks requires modeling of numerous satellites and their interactions.

    \item \textbf{Responsiveness:} As the network grows and the simulation becomes more computationally demanding, there is a risk that the simulator could slow-down, affecting its ability to run in real-time or near-real-time. This is particularly concerning for testing production-grade applications, where any discrepancy in timing (clocking) between the simulation and real-world operation can lead to inaccurate results. Slow simulations also make non-urgent simulations less agile, since it is common to rerun simulations multiple times for statistical validity, or with slightly changed assumptions to test different ``what if'' scenarios. We desire a simulator which runs in real time, i.e., 1-second of simulation requires 1-wall clock second.
    
    \item \textbf{Fidelity:} The cornerstone of any simulator is its ability to realistically mimic the real-world environment of the mega-constellation satellite networks. This encompasses accurately simulating link characteristics, antenna and RF parameters, and the impact of weather conditions. Furthermore, having the interfaces to allow incorporating production-grade transport and application protocols is essential for high-fidelity simulations. 

    \item \textbf{Low Footprint} Finally, we require that the simulator can achieve its functionality with low computational footprint (in terms of CPU and memory usage, as well as in terms of the number of machines or virtual machines required.).

\end{enumerate}

\begin{figure*} %h
    \centering
    \includegraphics[width=0.99\linewidth]{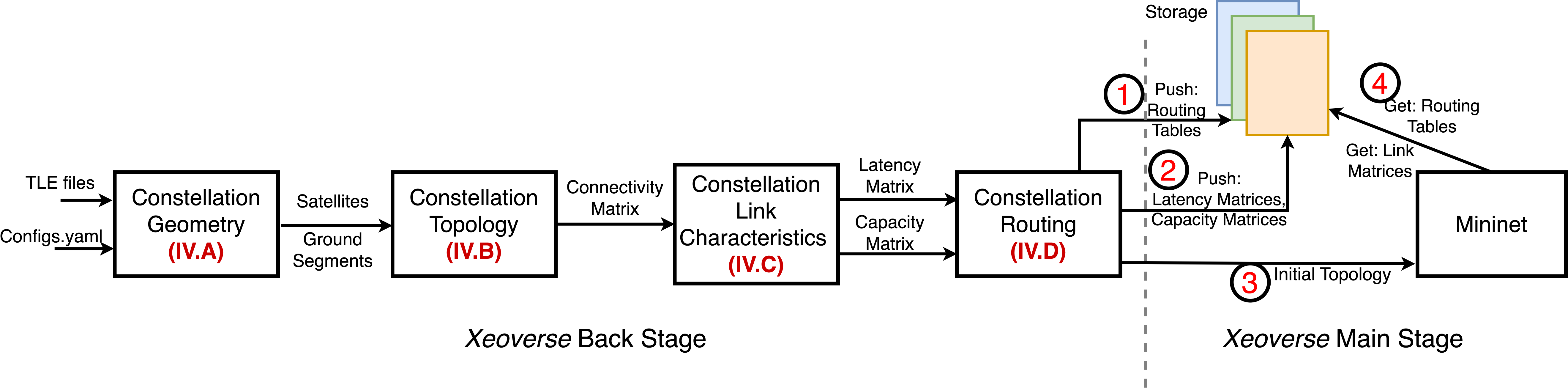}
 \vspace{-.1in} 
\caption{Workflow of \simulator Simulator. \textmd{The process is divided into two stages. The \simulator Back Stage builds the constellation topology, calculate the link characteristics in terms of latency and capacity, and finally compute the routing tables for each satellite. The \simulator Main Stage configures a Mininet topology based on the inputs from \simulator Back Stage, and run applications on top of Mininet.}}
\vspace{-.2in}
\label{fig:architecture}
\end{figure*}
\vspace{-.1in}
\subsection{Previous approaches}
There are different design approaches to build network simulators for LEO satellite mega-constellations. We review two prominent approaches, and compare them with our desiderata.

\textbf{Hypatia} is a network simulator designed for LEO networks, as an add-on module on top of the well-known NS-3 simulator. The choice to build upon NS-3 offers significant advantages, particularly in the ease of developing and testing new transport and application layer protocols for LEO networks, given the extensive tooling offered by NS-3 as well as implementations of several major protocols, including experimental proposals from various research papers. Furthermore, NS-3 protocols are implemented in user-space, which is significantly simpler than in-kernel implementations. 

However, this approach is not without its limitations. The reliance on NS-3, which is a discrete event network simulator, introduces challenges in meeting the above design requirements of scalability and responsiveness. The primary issue stems from the large volume of events Hypatia must process in a short time to accurately simulate the interactions and behaviors of dynamic LEO networks. This event-driven simulation model, while detailed, poses scalability (\textbf{D1}) and responsiveness (\textbf{D2}) challenges due to the computational demands of processing a large number of events, especially in scenarios featuring high traffic volumes and large number of nodes as highlighted in \S\ref{sec:evaluation}.

\textbf{StarryNet} adopted a different approach to simulate LEO mega-constellation networks, utilizing  Docker container technology to create this simulator. Each network element, including user terminals, ground stations and satellites, is represented by a Docker container, with virtual network interfaces established between them to simulate ISLs and GSLs. 

This overcomes  some of the scalability problems of the discrete event simulation approach of Hypatia, but there are still certain challenges, especially regarding scalability, responsiveness and footprint: The Docker's bridge interface, which is essential for StarryNet's simulation to emulate the ISLs and GSLs, imposes a limit on the number of containers that can be attached to a single virtual bridge interface. Specifically, the maximum is capped at 1023 containers\footnote{https://docs.docker.com/network/drivers/bridge/, this is due to the limitation set by the Linux kernel bridge interfaces to allow a max of 1024 ports.}, effectively restricting the simulator to representing no more than 1023 ground segments plus satellites on a single machine (impacting \textbf{D1}). To address larger constellations, StarryNet employs distributing the simulation across multiple machines, increasing setup complexity due to the intricate configuration of inter-container links, and affecting \textbf{D4}, the footprint required for the simulation. Furthermore, StarryNet's choice of implementation introduces performance bottlenecks. In particular, the way in which StarryNet employs threading encounters issues with the Global Interpreter Lock (GIL) in Python (detailed in \S\ref{sec:evaluation}). StarryNet spawns one thread per link to update these link characteristics, resulting in a total number of threads equal to the number of links in the megaconstellation. In \S\ref{sec:evaluation} we show how this mechanism, which prevents multiple native threads from executing  at the same time, significantly slows down the StarryNet simulation as the size of the constellation increases.

\vspace{-.12in}
\subsection{xeoverse Overview}

\begin{figure} %h
    \centering
    \includegraphics[width=0.65\linewidth]{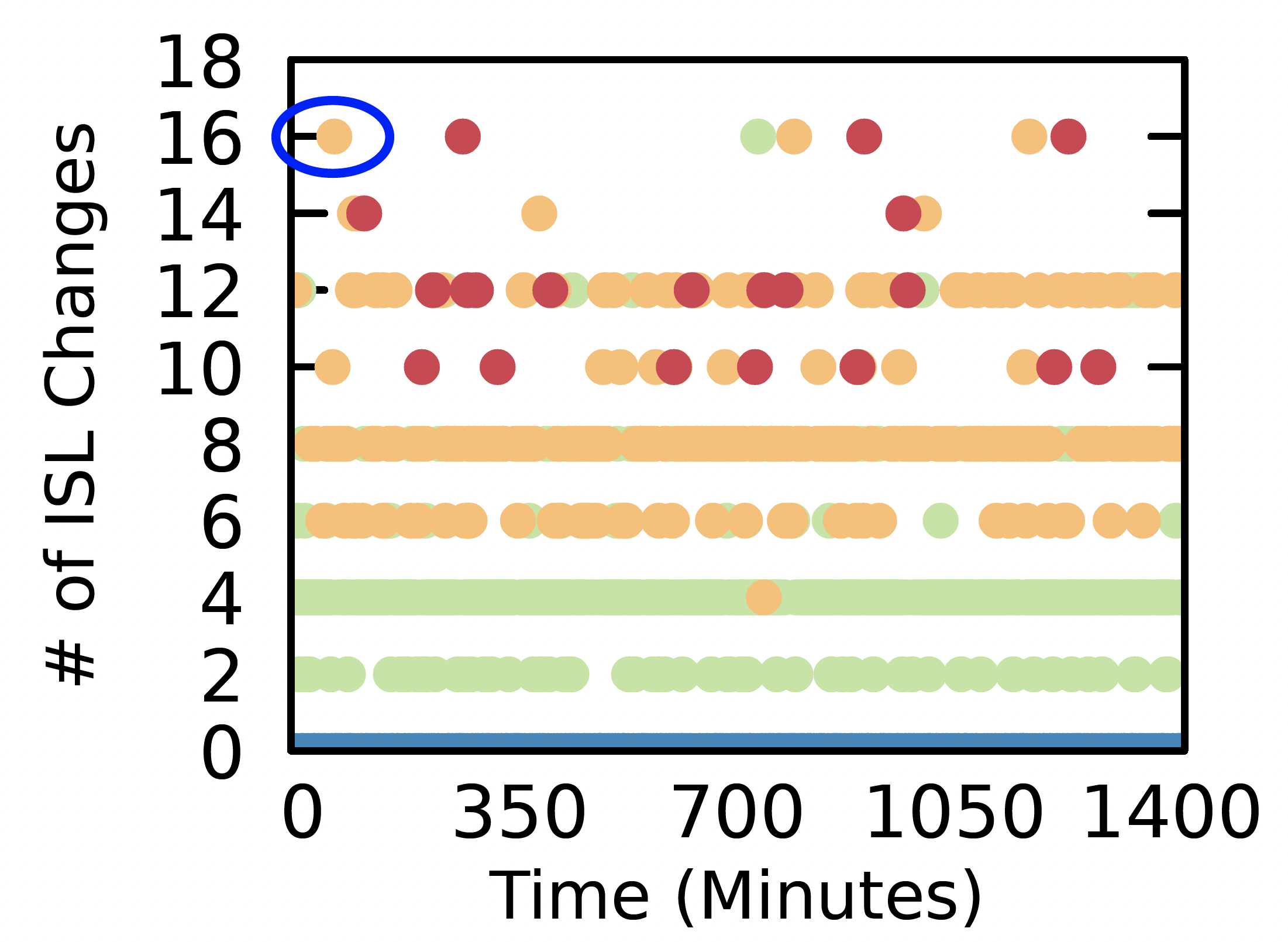}
 % \vspace{-.1in} 
\caption{LEO Networks topology changes. \textmd{Number of ISLs changes (left y-axis) per minute using colored dots: red for three occurrences, orange for two, and green for a single occurrence within each minute, visually representing the frequency and distribution of ISL changes.}}
\vspace{-.25in}
\label{fig:isl_changes}
\end{figure}

% \vspace{-.05in}
Fig.\ref{fig:architecture} presents a high-level schematic of \simulator, developed to satisfy the above desiderata. With scalability (\textbf{D1}) in mind, we based \simulator on Mininet, a network emulation platform that relies on  lightweight process-level virtualization technology to emulate thousands (up to 4096) of network nodes on commodity machines, allowing a small footprint for the entire constellation (\textbf{D4}). 

Mininet represents each network node (or host) as a lightweight virtual process (\textbf{D4}) that is connected to one or many virtual interfaces using a Linux Kernel feature called \textit{network namespace}. This feature allows each network node to have isolated network environments and routing tables. 

The fact that Mininet is developed over the standard Linux Kernel and each network node is a virtualized process allows many Linux-based network testing applications, e.g., iperf, ping, and traceroute., to be easily deployed on these nodes, helping fidelity (\textbf{D3}). This also facilitates testing the performance of the state-of-the-art network/application/transport-layer protocols (e.g., BBR, CUBIC, HTTP/3) that have been already developed or have reference implementations in standard Linux (\textbf{D3}). In addition, \simulator achieves better fidelity \textbf{(D3)} by modeling link characteristics to include both ground segment and satellites RF parameters and other influencing factors, such as weather conditions, which are not addressed by Hypatia and StarryNet. 

Each satellite, user terminal and gateway is represented as a separate Mininet node. These are connected together to form the global satellite constellation for the simulation scenario under consideration. The challenge lies in the fact that there are thousands of nodes in LEO mega constellations, and connections are made and broken as the satellites move over the user terminals and gateways on earth, creating constant topology changes.
Our approach to responsively (\textbf{D2}) managing this link churn inherent to LEO mega constellations is informed by three key observations:

\noindent\textbf{I. Satellite movements are predictable.}  Satellites orbit the Earth in predefined paths (orbits) at constant speeds.  This predictability allows us to pre-compute, for any given time point, the position of all satellites in a constellations, their link characteristics, as well as the entire network topology  -- inter-connectivity among satellites. This significantly reduces the need for real-time calculations during the simulation.

Furthermore, we do this pre-computation phase only once for a given scenario, and then the main simulation uses this pre-computed data to rerun multiple times. This approach is particularly beneficial for various purposes e.g., ensuring statistical validity, testing different variants of the same protocol, or conducting comparisons of a new proposal against various baseline implementations. For instance, a new congestion control protocol can be compared against common TCP variants, CUBIC, BBR, etc., within the same simulation scenario, with each protocol tested using data generated by a one-time pre-computation run. By separating the simulation into a pre-computation phase and the main simulation, we significantly decrease the overall time required for the simulation. Statistical validity may require that the same experiment be run multiple times, which again benefits from reusing the pre-compute phase across the different runs.

\noindent\textbf{II. Only few Inter-Satellite Links change over short time periods.} Despite the fast mobility of LEO satellites -- approximately 27,000 kilometers per hour -- the frequency of link changes is relatively low. This observation is demonstrated in Fig.\ref{fig:isl_changes}, which depicts the number of ISL changes (on y-axis) per minute over the entire Starlink \textit{shell 1,} -- consisting of $1584$ satellites, over a span of $25$ hours, depicted on the x axis as minute 0 to minute 1500 (25*60). We use the grid topology to build the ISL links, and the total number of ISLs is 5,346 links. The figure uses colored dots to indicate the distribution of these ISL changes within each one-minute interval. Specifically, a red dot signifies that the ISL changes occurred in three separate instances within the minute, an orange dot represents ISL changes happening twice within the minute, and a green dot indicates a single occurrence of ISL changes in the minute. For example, in the upper-left part of the figure (highlighted with the blue circle), orange dots reflect $16$ ISL changes per minute (, i.e., a mere 0.3\% of the 5346 total ISL links), with these changes evenly split into two instances (hence orange), each comprising $8$ ISL changes within the one-minute window. This observation suggests that it is unnecessary to update all satellite links continuously; instead, focusing on a small subset of changing links can substantially reduce computational demands. 

\noindent\textbf{III. Many links may not matter in a  simulation scenario} The necessity to update links can be further minimized by concentrating solely on those links that are crucial to the simulation's current focus -- namely, links that connect endpoints involved in the simulated scenario. For example, if a simulation runs flows on the North-South axis between  end hosts in London and Johannesburg, with East-West cross traffic running between different European cities (e.g., Brest, France to Berlin, Germany; Zagreb, Croatia to Zaragoza, Spain, etc.), parts of the constellation over Asia or the Americas may not need to be simulated. By prioritizing updates to links of interest, we can further minimize  computational demands.

Driven by Observation \textbf{I}, \simulator is architecturally divided into two primary components: \textbf{\simulatorPython} and \textbf{\simulatorMininet}~\footnote{This nomenclature draws on the metaphor of a drama theatre's backstage and main stage dynamics, where the backstage activities prepare for the final performance on the main stage.} \simulatorPython~is a set of modules designed to leverage the predictable nature of LEO satellites. This segment of the \simulator is responsible for the preliminary computations of the mega-constellation's topology, routing paths, and link characteristics. By capitalizing on the predictability of satellite movements, \simulatorPython~enables \simulator to perform simulations in real-time. The output of these modules is then fed to the \simulatorMininet~ (i.e., the Mininet emulation) to play the emulated scenario, and run the required applications. 

Driven by Observations \textbf{II} \& \textbf{III}, \simulatorPython pre-computes the topology changes on a second-by-second basis\footnote{This interval for updating the network topology is configurable.} during the whole simulation timeframe. This information is then used to manage in real-time the connectivity between different nodes of all kinds (satellites, user terminals and gateways), rewiring only those links that have changed in the past second. Collectively, \textbf{I}, \textbf{II} and \textbf{III} thus help ensure \textbf{D2}.

\vspace{-.03in}
\section{Design and Implementation}
\label{sec:implementation}
% \vspace{+0.15in}

We now discuss the detailed design of \simulator, explaining the functionality of every module, walking through the \simulator pipeline (Fig.~\ref{fig:architecture}).

\vspace{-0.05in}
\subsection{Constellation Geometry}
% \vspace{+0.15in}
As depicted in Fig.\ref{fig:architecture}, \simulator starts with the configuration file which is a \textit{.yaml} file divided into $7$ sections that configures constellation parameters, satellite positions, ground station locations, RF parameters, simulation time, weather conditions, and applications to be tested on the simulated constellation. We use the Two-Line Element (TLE) format to define the satellites’ positions and trajectories for a given period of time. The TLE format contains all the orbital parameters (i.e., Keplerian elements) required to propagate the satellite using different orbit propagator techniques such as J2, J4, and SGP4. For operational satellites, the TLE data is available on Celestrak~\cite{celestark}. However, this format is insufficient to study the network adjacency or the connectivity between different satellites or between a satellite and a ground station. To understand the adjacency of this large-scale network, we need to know which satellites belong to which orbits and what are the adjacent orbits. To achieve that, we built a \textit{utility} that accepts the TLE format as input and outputs the list of orbits and the list of satellites within each orbit (i.e., a list of list). The list of satellites within one orbit is sorted according to the calculated satellite \textbf{\textit{true anomaly}} value. The true anomaly is the angle between the direction of periapsis and the current position of the satellite. The true anomaly is calculated based on the eccentricity ($e$) -- the first orbital parameter in the TLE format --, mean anomaly ($M$) -- the last orbital parameter in the TLE format -- and eccentric anomaly ($E$). We derive the true anomaly ($\nu$) using the following equations\cite{battin1999introduction}:
\vspace{-.07in}
\begin{gather*}
 M = E - e \sin E \quad \cos\nu = \frac{\cos E - e}{1 - e\cos E}
\end{gather*}

\vspace{-.2in}
\subsection{Constellation Topology} Once the orbits and satellites within each orbits are identified, \simulator starts to build the constellation topology. \simulator implements two types of topology: the bent-pipe, and the ISL topology. In the \textit{bent-pipe topology}, the traffic travels from the ground station to an available satellite overhead and goes back again to Earth through the closest visible gateway. The traffic then continues to the destination through emulated fiber connectivity. In the \textit{ISL topology}, the traffic can potentially traverse multiple hops of satellites through ISLs until it returns to the ground terminating at the destination. 

As proof-of-concept, \simulator implements two patterns of ISL connectivity. The first pattern simulates ISLs between satellites within the same orbit, establishing two links per satellite (one link to the satellite preceding it, and another to the satellite following it in the orbit).  The second pattern implements the Grid ISL topology where each satellite establishes 4 ISLs to the adjacent satellites, two to the satellites in the same orbit, and two to the satellites in the two adjacent orbits. We choose to implement this ISL connectivity pattern based on the literature and recent FCC filing documents which indicate this is used by SpaceX's Starlink, Amazon Kuiper~\cite{fcc_2021, sp_2020, kuiperfcc, starlink-fcc1}. However, adding other connectivity patterns is as simple as adding one function to the constellation geometry module. 

The Ground-to-Satellite Links (GSLs) of the constellation are defined based on how terminals decide on switching from one satellite to another, which is also known as the \textit{handover strategy}. \simulator currently implements two handover strategies which represent the upper and lower bound. \textbf{Distance-based} is the handover strategy that assumes the terminal will be attached to the nearest satellite (i.e., minimum distance). Hence, terminals periodically assess the distance to all visible satellite and remain or switch to the nearest satellite when needed. The second handover strategy is the \textbf{Longest-attachment} where the terminal attaches to one satellite and will remain attached to it until it disappears from view, before switching to another satellite. 

The output of this constellation topology module is a $N \times N$ matrix where $N$ is number of network nodes (i.e., satellites, terminals, and ground stations) where $N[i][j] = 1$ if a link can be established between node $i$ and $j$, and zero otherwise. We name that matrix, \textit{\textbf{the connectivity matrix}}. Each simulated instant has its own connectivity matrix; hence the matrix changes over time. 
~\vspace{-0.03in}
\subsection{Constellation Link Characteristics}
As illustrated in Fig.\ref{fig:architecture}, the third block in the pipeline is \textit{Constellation Link Characteristics} where \simulator assigns (i) latency, (ii) capacity, and (iii) signal-to-noise ratio (SNR) as three attributes for both Ground Satellite Links (GSLs), and Inter-Satellite Links (ISLs). The latency attribute is calculated as a propagation delay between the two endpoints of the link. The SNR of the GSLs is calculated as a function of configurable RF parameters (e.g., TX power, receiver sensitivity, channel width, etc)\footnote{In \S\ref{sec:evaluation}, we used the RF parameters extracted from FCC filing for Starlink}, and different losses variables such as losses due to weather conditions, antenna polarization, and mis-alignment. Our channel modelling is based on on the ITU-T model for Ku- and Ka-band\cite{iturpy-2017}. Specifically, we defined the SNR as: 
$PW_{i}^{EIRP} - BW - FSPL - L - G/T - k $ where $PW_{i}^{EIRP}$ is the Equivalent Isotropic Radiated Power (EIRP) of a node, $BW$ is the channel bandwidth, $FSPL$ is the free-path loss model, $L$ is the total losses due to antenna polarization and mis-alignment, in addition to the attenuation due to weather conditions. We leverage Open Weather Map API \cite{weatherapi} to get  fine-grained weather information per location (i.e., lat/long) such as rain/snow levels, temperature, humidity and pressure. $G/T$ is antenna gain-to-noise-temperature ratio, and $k$ is the minimum sensitivity of the receiver (i.e., Boltzmann’s constant). The capacity of the GSLs is defined as a function of SNR, bandwidth and cell density and using the classical Shannon capacity equation. Cell density is a configurable parameter defined as $1/\mathit{Number-of-Users-per-Cell}$, which is then used to define the fair-share of cell capacity. The cell density can be tuned in the \simulator configuration file. While the GSL capacity is calculated based on SNR, bandwidth and cell density which vary over time, the ISLs capacity is set with a fixed configurable value due to the optical nature of ISLs, and the fact that these links are less vulnerable to losses and interference as they operate in outer space, beyond the earth's atmosphere. The output of this module is two matrices, $LM$ and $CM$, representing a latency matrix and capacity matrix, respectively. Both matrices are of size $N \times N$, where $LM[i][j]=x$ indicates that the latency between nodes $i$ and $j$ equals $x$ milliseconds. Similarly, $CM[i][j]=y$ indicates that the capacity between nodes $i$ and $j$ equals $y$ Mbps.
\vspace{-.03in}
\subsection{Constellation Routing}
\simulator supports static routing protocols. The link characteristics matrices (i.e., latency and capacity) are passed to the constellation routing module to build the routing tables for each satellite in the constellation. \simulator uses the Python library \textit{networkx} to map these matrices to a network weighted graph where the edge weight can be set according to the latency or the capacity. We then use \textit{Dijkstra} algorithm to calculate the shortest path between any two nodes on the graph. The constellation routing module can also be configured to compute the shortest routes based on hop counts by setting the weights of the graph to $1$. The constellation routing module then converts these shortest-path routes to Linux-based \textit{ip route} commands that are saved in timestamped files fetched later by the Mininet module to run the simulation.

\vspace{-.03in}
\subsection{Putting it all together}

\simulator initializes a Mininet topology using the specific configurations provided by the constellation routing module, including latency and capacity matrices, along with Linux routing commands for setup in the Mininet virtual machines (VMs). It is important to note that the constellation routing module selectively passes information on only the relevant satellites and ground segments (i.e., satellites, ground segments that connect endpoints involved in the simulated scenario), along with their link characteristics, rather than the entire mega-constellation. \simulator then spins out a background process to keep track of the time and update the Mininet topology and VMs routing at every configurable interval, to match the current connectivity between the moving LEO satellites and the ground stations and user terminals currently under them. The default interval for topology updates is $1$ second but this can be configured. \simulator users can employ standard Linux-based tools such as \textit{iperf}, \textit{ping}, and \textit{traceroute} to test latency, throughput etc., across scenarios of interest, such as different congestion control algorithms, or routing and path changes between any two endpoints. 
\vspace{-.05in}
\section{Evaluation}
\label{sec:evaluation}

In this section, we evaluate the performance of \simulator in two stages. Our aim is to rigorously assess its capabilities in comparison to Hypatia and StarryNet, and to demonstrate its accuracy in replicating real-world network conditions. We initiate our evaluation by assessing the scalability of \simulator. Here, our focus is to ascertain \simulator's capability in managing large-scale simulations by comparing \simulator with Hypatia and StarryNet and  aim to understand how it performs in handling extensive network models. In the second stage, we evaluate the fidelity of \simulator, segmented into two analyses: Initially, we assess how \simulator simulates the performance of congestion control over ISL paths between different ground segments. We also highlight that such impact of ISLs changes are often overlooked by other simulators. Subsequently, we compare the \simulator's performance against the real-world network performance of Starlink. This comparison is particularly focused on \simulator's accuracy in reflecting network behavior under diverse weather conditions, emphasizing its high fidelity in replicating real-world  scenarios. 

\begin{figure}[t]
  \centering
 \includegraphics[width=0.85\linewidth]{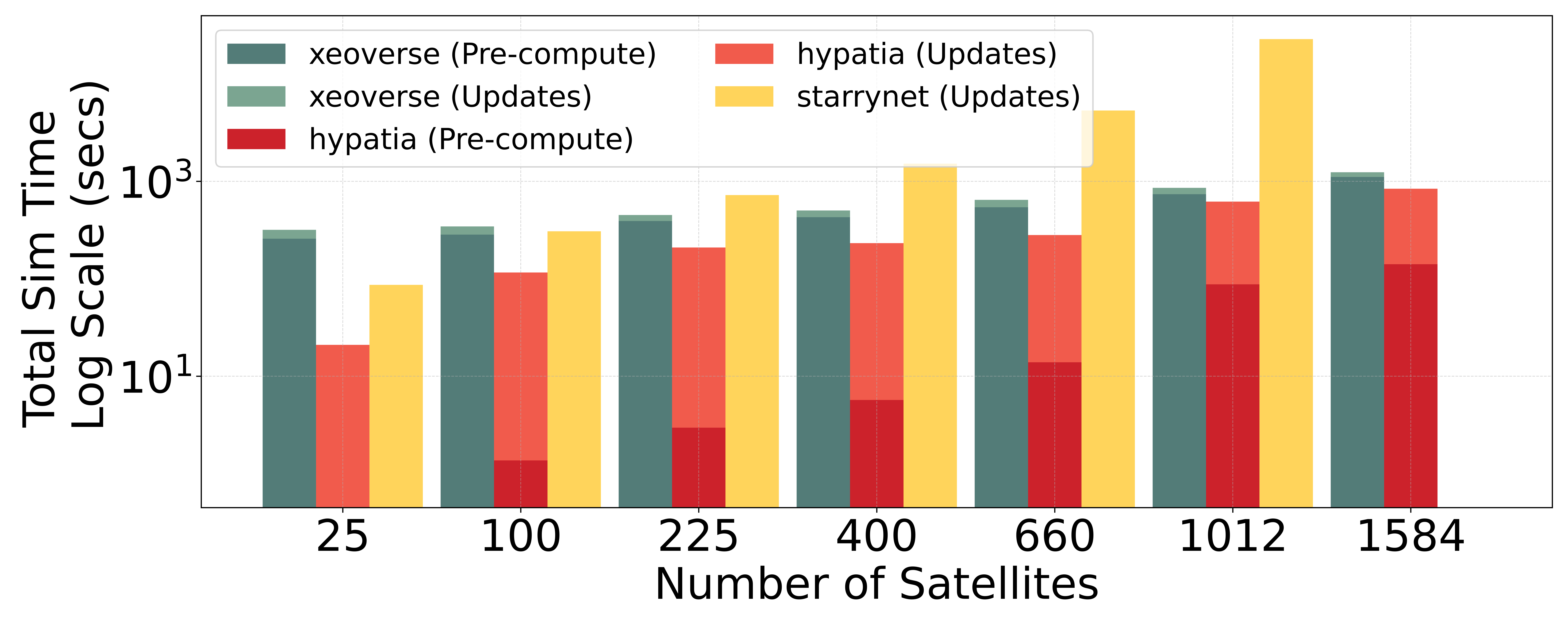}
 \vspace{-.05in}
\caption{\textmd{The total simulation time (including pre-compute and simulation updates times) for the 3 simulators when varying the number of satellites from $25$ to $1584$. \simulator's simulation update time is $2.3$x and $40$x times less than Hypatia and StarryNet, respectively.}}
\vspace{-.25in}
\label{fig:scalabilitya}
\end{figure}

\vspace{-.02in}
\subsection{Scalability and Responsiveness}

We compare the scalability of \simulator against Hyaptia and StarryNet simulators by varying two parameters: the number of Starlink satellites used and the number of ground stations. This approach enables us to understand how each simulator copes with increasing network complexity. To ensure fair comparison, we perform our analyses on a high-performance setup featuring $26$ cores and $64$ GB RAM.

Fig.\ref{fig:scalabilitya} highlights the simulation times required by Hypatia, StarryNet, and \simulator for varying sizes of satellite constellations, ranging from a small group of $25$ satellites to a full constellation of $1584$ satellites --  mirroring the configuration of Starlink's \textit{Shell 1} which consists of $72$ orbits with $22$ satellites each. Within this context, \textit{total simulation time} includes both the pre-computation and simulation update phases. Pre-computation involves calculating the LEO network's state ahead of the simulation, covering aspects such as ISLs, GSLs, their characteristics, and network routes. For StarryNet, there is no pre-computation phase; all the calculations for the ISL and GSL changes are done online. Therefore, for StarryNet, the total simulation time spent is the simulation update phase.

The simulation update phase is an ongoing process that updates the network topology (and inter-connectivity) at pre-defined intervals (e.g., every 100ms, 1s or 10s), incorporating changes in link latencies, the creation or deletion of links, and modifications to link capacities. As the constellation size increases, StarryNet's simulation time grows exponentially, particularly during the update phase. For instance, simulating one second of real-time network operation can take up to $79$ seconds when simulating a constellation of $660$ satellites ($30$ orbits, $22$ satellite per orbit). Please note that Fig.~\ref{fig:scalabilitya} is presented in  log-scale for better readability. This inefficiency primarily arises from StarryNet's architecture, which generates a proportionate number of threads to the number of links, leading to substantial delays. For a constellation of $660$ satellites, StarryNet launches $1323$ threads to update the ISL and GSL links, encountering a bottleneck where the simulator spends extensive periods waiting for threads to execute sequentially due to Python's Global Interpreter Lock (GIL)\cite{gil}. This limitation prevents simultaneous thread execution, causing significant slowdowns in CPU bound processes such as the simulator threads for updating link topologies. Spawning multiple threads here (1 thread per link in the topology) incurs the overhead of thread management without the benefits of parallel processing. Furthermore, StarryNet's update mechanism, which updates link characteristics at fixed intervals regardless of actual changes in topology, contributes to its scalability issues. This approach leads to unnecessary computations, particularly as the constellation's size expands. 

\begin{figure}[t]
  \centering
 \includegraphics[width=0.67\linewidth]{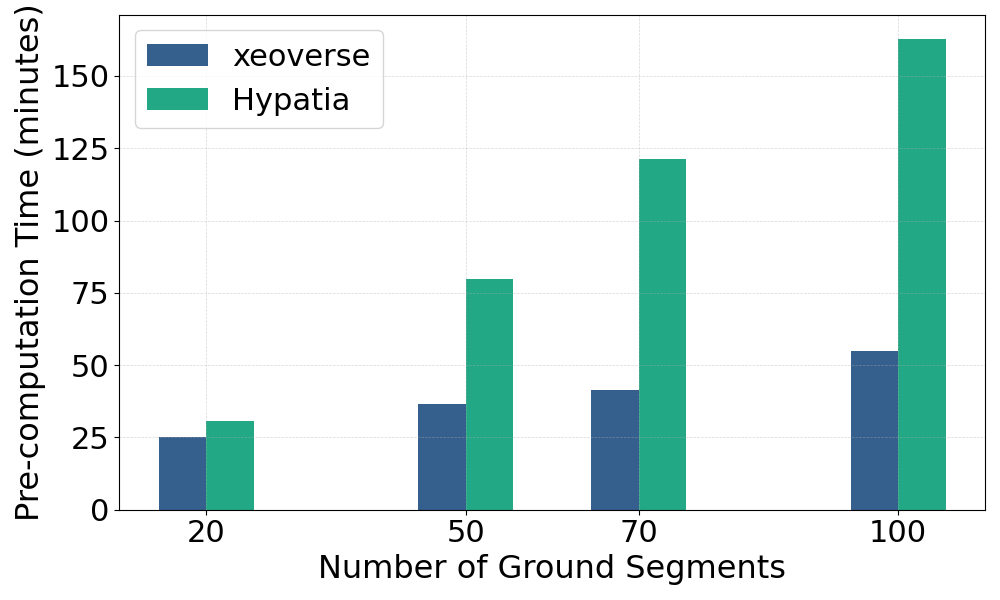}
 \vspace{-.08in}
\caption{\textmd{Focusing only on the pre-compute time due to its computational demanding nature, and varying the number of ground segments from $20$ to $100$. We also focus only on Hypatia and \simulator as we simulate a full constellation of $1584$ satellites in this experiment which cannot be achieved by StarryNet. \simulator's pre-computation time is $2.9$x times less than Hypatia.}}
\vspace{-.25in}
\label{fig:scalabilityb}
\end{figure}

As previously mentioned in \S\ref{sec:design}, StarryNet's use of Docker container technology and the inherent limitations of Docker's container bridge interface prevent it from simulating more than 1023 elements on a single machine. Hence, there is no data point in Fig.\ref{fig:scalabilitya} for StarryNet involving 1584 satellites.

\begin{figure*}[t] %h
  \centering
  \subfigure[]
  {
    \includegraphics[width=0.23\linewidth]{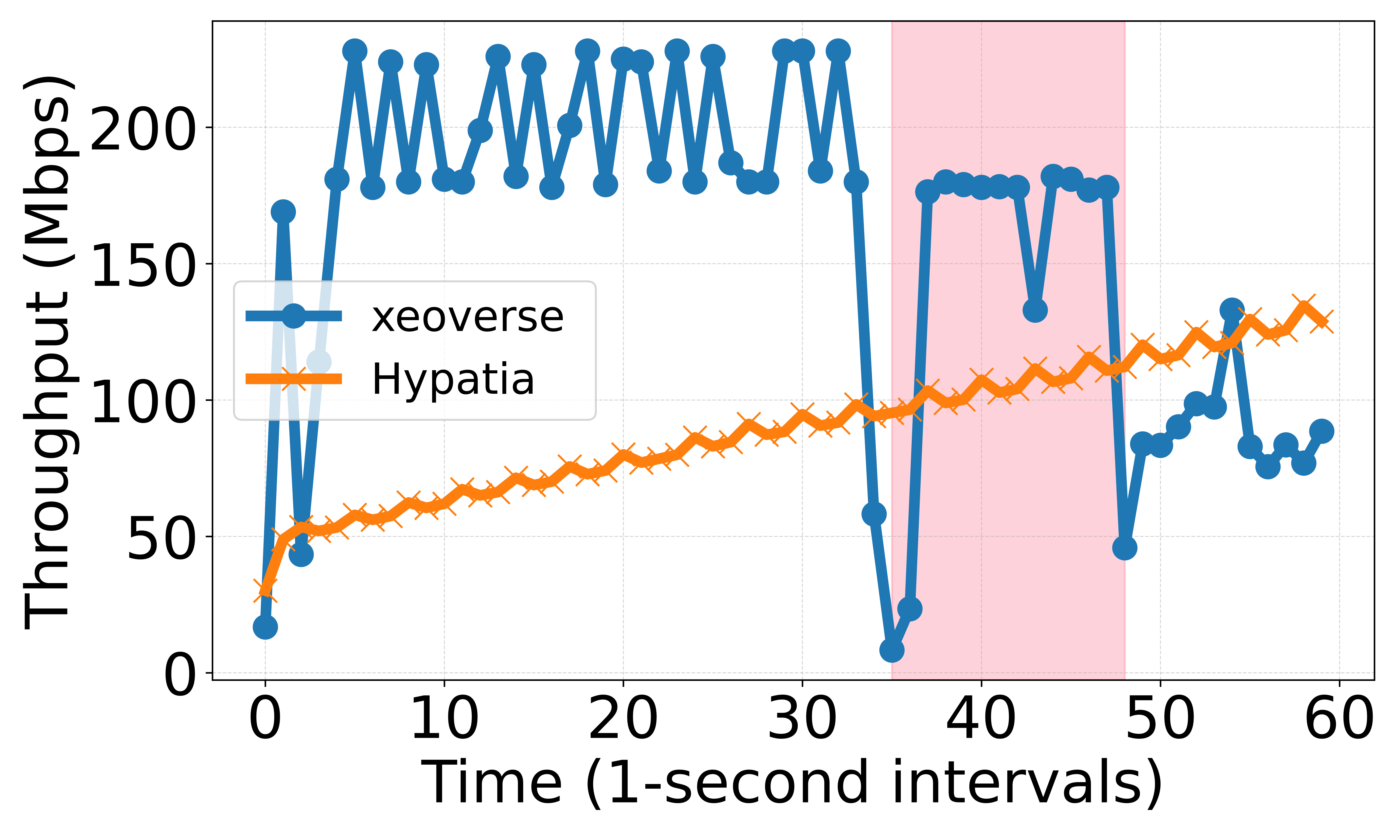}
    \label{fig:thruput-NC-LON-BC}
  }
%   \hspace{-.15in} 
  \subfigure[]
  {
    \includegraphics[width=0.23\linewidth]{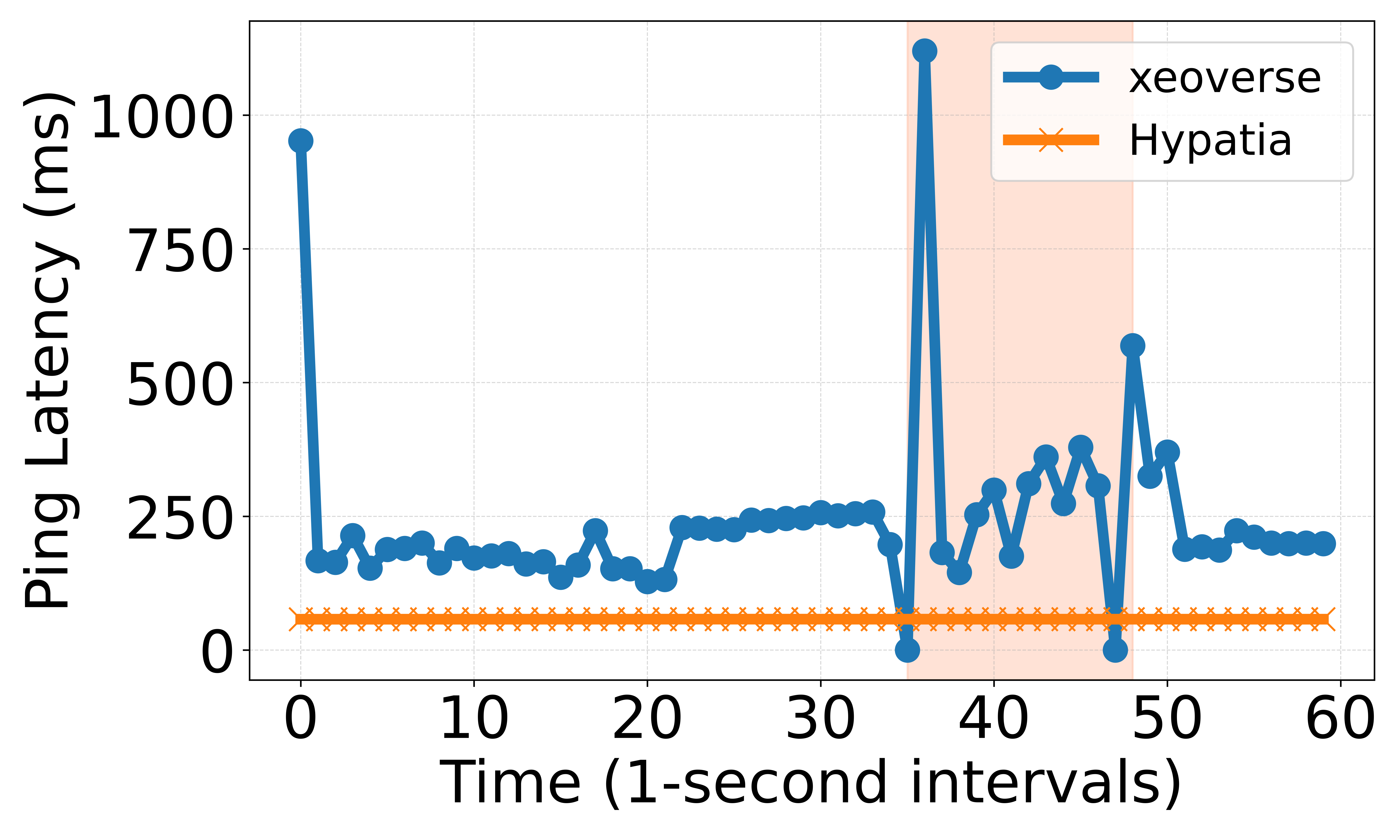}
    \label{fig:thruput-time}
  }
  \subfigure[]
  {
    \includegraphics[width=0.23\linewidth]{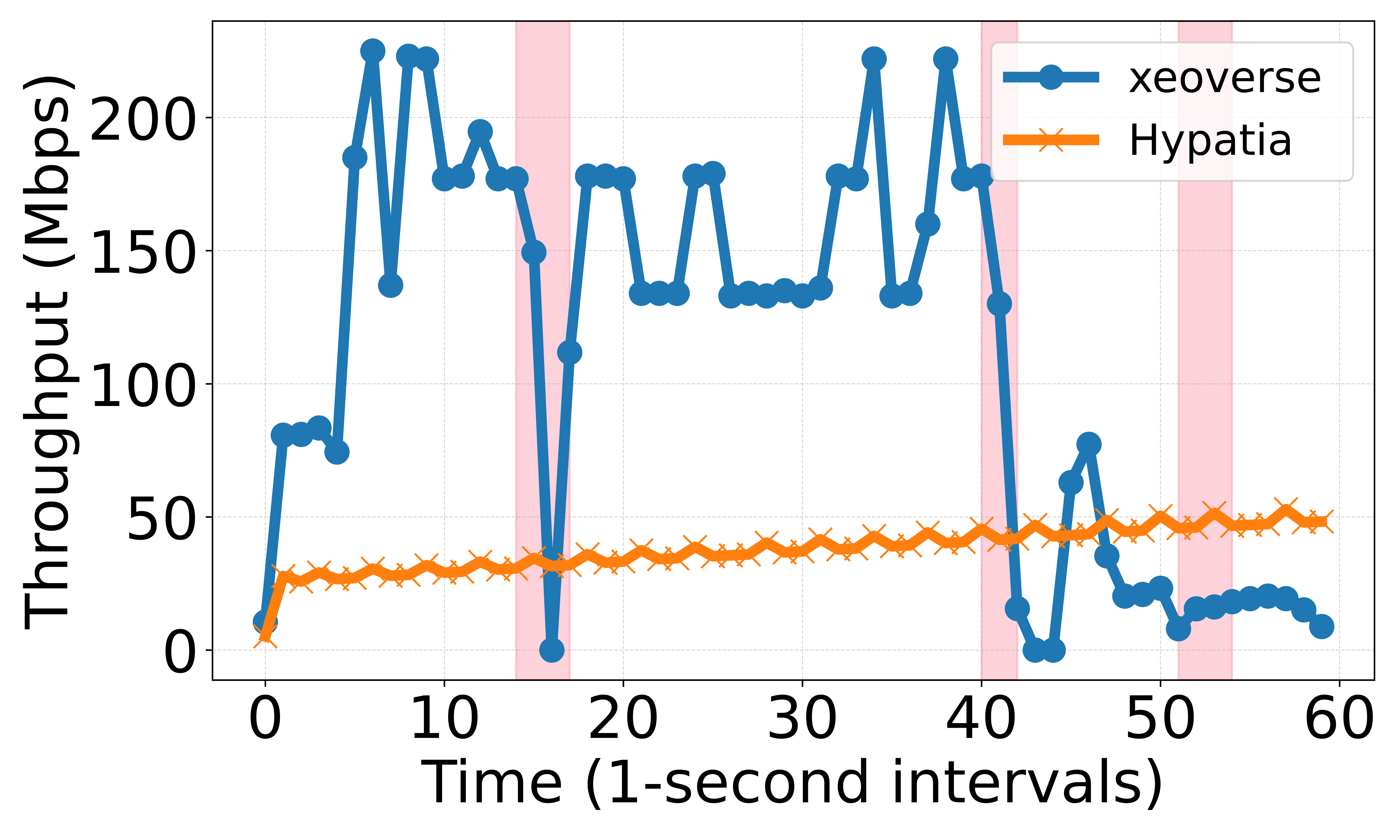}
    \label{fig:thruput-time}
  }
  \subfigure[]
  {
    \includegraphics[width=0.23\linewidth]{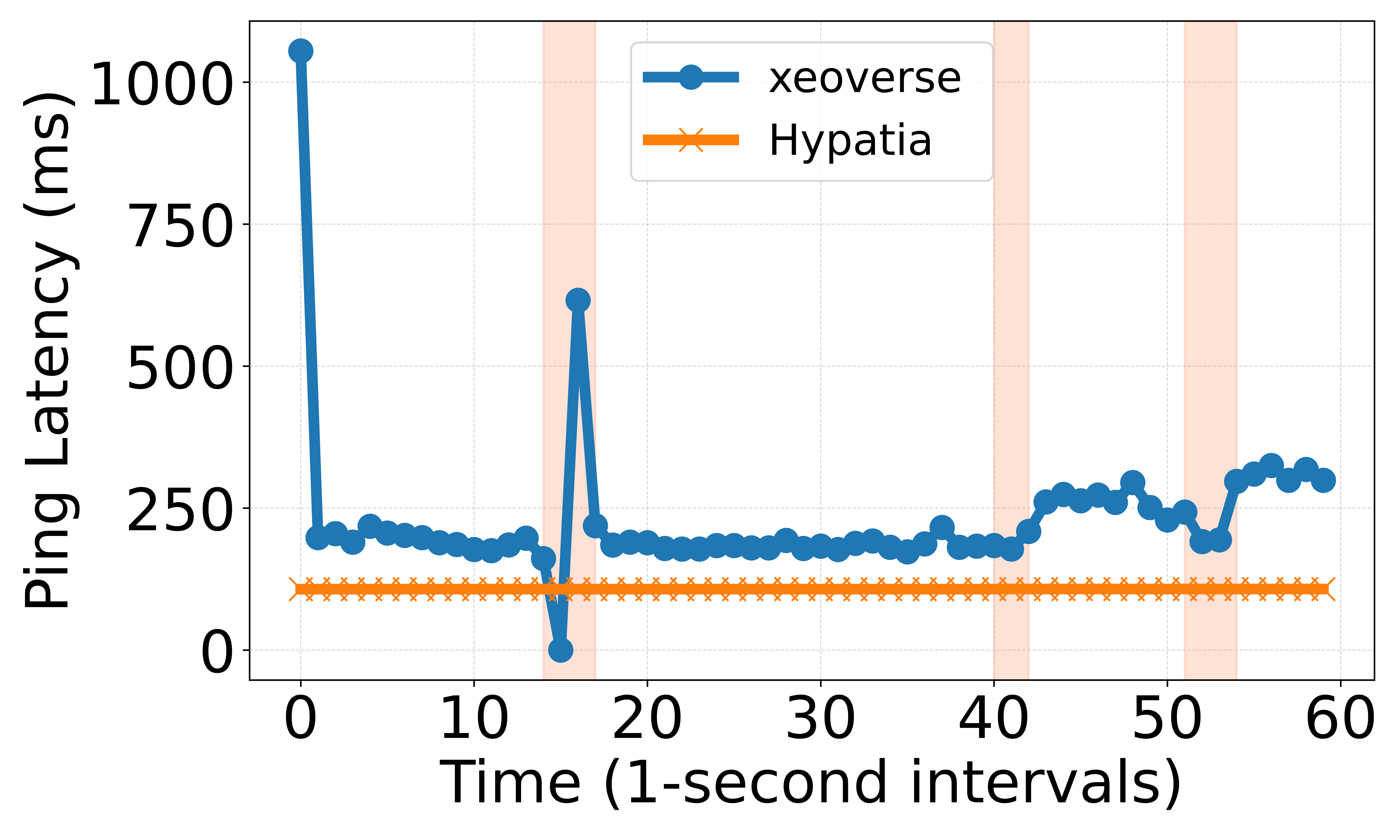}
    \label{fig:thruput-time}
  }

\vspace{-.10in}
\caption{\textmd{\simulator shows high-fidelity when compared to Hypatia. (a) shows how the changes in the ISL path between two terminals one in London and another in San Francisco impacts the observed data rate at these terminal (the shaded red area); yet that is not reflected in a replicated scenario in Hyptia. (b) shows the latency spike for the same scenario when the ISL path changes. (c) shows a rapid fluctuation of throughput for multiple path changes between two terminals in Paris and Sydney. (d) shows the latency spike between Paris and Sydney when the ISL path changes.}}

\vspace{-.28in}
\label{fig:fidelity1}
\end{figure*}

On the other hand, Hypatia performs very well in terms of pre-compute time due to its simplified, or low-fidelity, approach during this stage. Hypatia focuses solely on constructing a basic topology map that outlines the connectivity between satellites and establishes network routes. However, it stops short of calculating detailed link characteristics, such as actual capacity, use of RF parameters, and SNR calculations. By adopting a simplified model that assumes link capacity is fixed at a maximum configurable value, Hypatia significantly reduces the computational effort required to determine the actual dynamics of link performance; yet, this decision also limits the fidelity of the simulation. Despite its fast pre-compute time, Hypatia's update times during the simulation phase do not achieve real-time performance. For example, simulating a 60-second real-world time with a constellation of $660$ satellites requires $266.76$ seconds of simulation time in Hypatia. If the simulation involves the entire constellation of $1584$ satellites, the simulation time extends to $698$ seconds for the same 60-second real-world time. This lag is largely due to its reliance on NS-3. The need to sequentially process every network update as discrete events, each representing a change in the network's state, contributes to the extended simulation times observed in Hypatia.

In contrast, \simulator capitalizes on the predictable nature of satellite constellations. By pre-calculating the entire network's topology, \simulator significantly speeds-up the simulation process. Although using this approach increases pre-compute time, which is approximately $2$ times higher than that of Hypatia, the rationale behind \simulator's strategy is to front-load the computational effort to accurately model the network's initial state, thereby facilitating real-time simulation updates. When it comes to simulation updates, \simulator outperforms both Hypatia and StarryNet by achieving real-time simulation. This means that one second of wall-clock time correspond to one second of simulation time in \simulator, making it $2.3$ times faster than Hypatia and $40$ times more faster than StarryNet, which is slowed down by its many threads and GIL.

In Fig.\ref{fig:scalabilityb}, we focus on the pre-computation phase specifically for Hypatia and \simulator. We are not able to simulate StarryNet due to the fact that in these experiments, we seek to simulate the entire constellation of 1584 satellites on a single machine, which is beyond the scalability limits of StarryNet. We vary the number of ground segments from $20$ to $100$ and introduce traffic flows between each pair of segments, resulting in a range of $10$ to $50$ flows (each flow connects one pair of ground-based elements). As the number of flows increases, Hypatia's pre-computation time significantly rises, reaching $162$ minutes for simulations involving $100$ ground segments with $50$ traffic flows in total. In contrast, \simulator maintains a faster pre-computation time of $55$ minutes under the same conditions, highlighting its superior performance in handling multi-flow experiments.

\vspace{-.04in}
\subsection{Fidelity}
\begin{figure}
    \centering
    \includegraphics[width=0.63\linewidth]{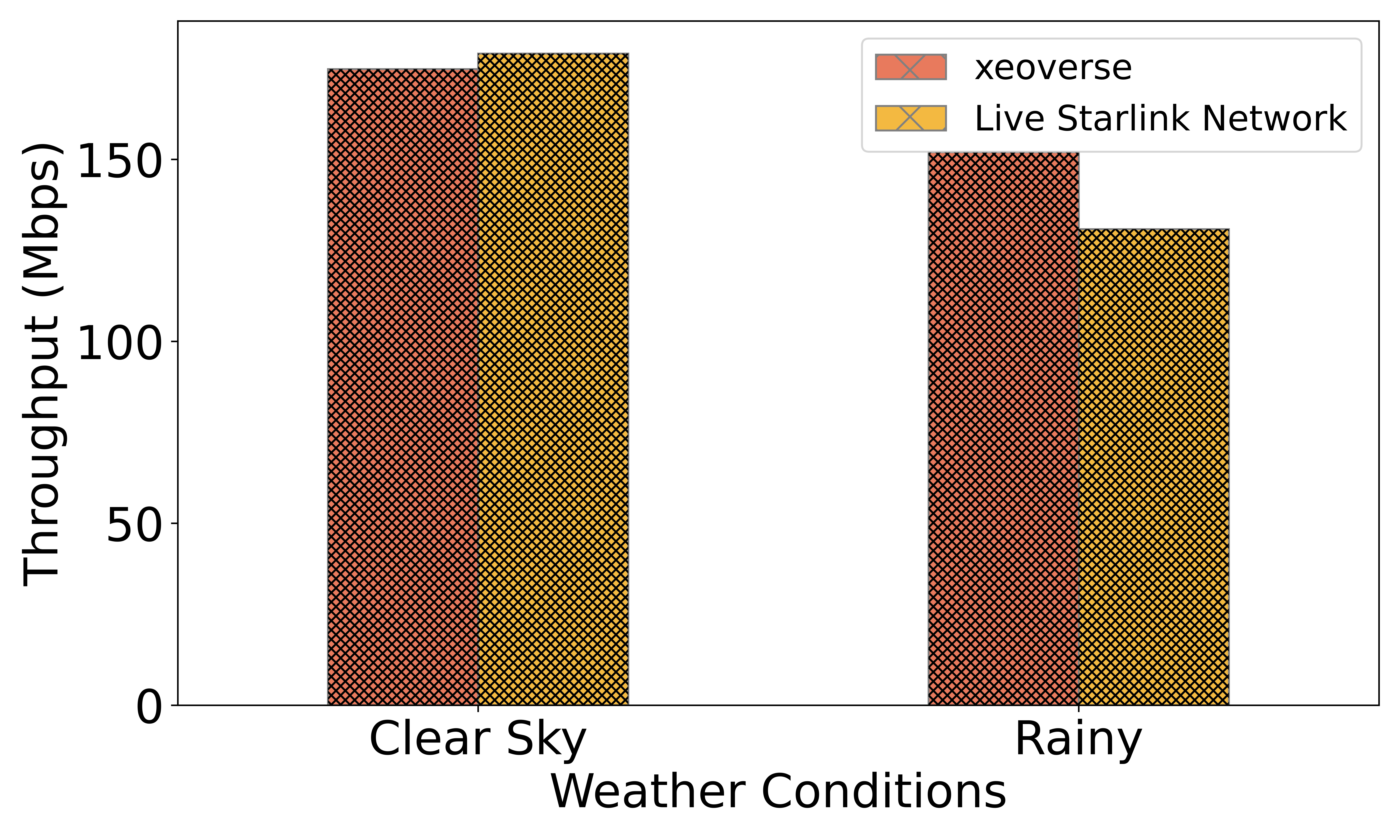}
    \caption{\textmd{\simulator captures the impact of weather conditions on the downlink throughput in the bent-pipe connection for Starlink user terminal. \simulator shows results that are close to reality: within 2\% for clear sky and within 16\% for rainy conditions.}}
    \vspace{-.25in}
    \label{fig:weather}
\end{figure}

\vspace{-.02in}
In our second series of experiments, we evaluated the fidelity of \simulator by examining how changes in ISL links affect TCP flow throughput and latency between ground segments in different locations, comparing the results with those obtained from Hypatia (we omit StarryNet as it is unable to scale beyond $\approx$ 1012 satellites which is insufficient for ISL-based East-West traffic to consistently flow over time, between the following cities). Specifically, we analyze traffic between two pairs of user terminals in London and San Francisco, and Paris and Sydney, over a 60-second duration for both TCP flow and ping tests.

Figures \ref{fig:fidelity1}(a)--(d) illustrate the outcomes for throughput and latency experiments for these two pairs, respectively. Fig.\ref{fig:fidelity1}(a), focusing on the connection between London and San Francisco, shows throughput variation occurs between $35$ and $48$ seconds. This fluctuation is attributed to a change in the ISL path connecting the two cities. During this period, the Starlink terminal in London switched its satellite connection from $STARLINK-2159$ to $STARLINK-30533$. As a result, the network path between London and San Francisco changed, increasing from an $18$ hop count to $21$. This change in the network route is also evident in the ping latency, where we notice a spike at the start of this interval. The mean latency surged by $67.6\%$ from the $35$-sec mark to the $48$-sec mark, compared to the latency observed before the path change. 

We conducted the same experiment for the second pair of user terminals between Paris and Sydney and observed similar performance patterns. Specifically, we observe ISL path changes at the $16$, $42$, and $51$-sec marks. These adjustments in the ISL paths directly impacted the connection between Paris and Sydney. For instance, at the $51$-sec mark, the throughput experienced a significant decrease, dropping from 77.3 to 8 Mbps. This experiment illustrates how dynamic ISL adjustments can profoundly affect network performance, mirroring observations made in London-San Francisco scenario.

Unlike the results observed with \simulator, Hypatia does not exhibit the same behavior regarding the impact of ISL changes on TCP flow throughput and latency. This discrepancy can be attributed to two main reasons. First, Hypatia operates as a module within NS-3, relying on the TCP and ICMP (ping) protocol implementations available in NS-3. While these implementations aim to mirror their real-world counterparts, there are differences that might not fully capture the details of actual network behavior. In contrast, \simulator can use the standard TCP and ICMP implementation. Second, in real network systems, a path change necessitates updates to the routing tables of the nodes involved in the changed path. This update process can lead to temporary packet drops or buffering as the new path information is propagated throughout the network. Such dynamics, intrinsic to real-world networking, are not mirrored in Hypatia's simulation environment. Unlike \simulator, Hypatia abstracts away the complexity of routing table updates following a path change, using abstracted APIs for node connectivity instead of simulating the detailed process of re-configuring routes.

In the second set of experiments to evaluate \simulator's fidelity, we focus on understanding how weather conditions affect downlink throughput in a bent-pipe architecture, specifically between a user terminal in our university campus and its associated gateway. We compare these simulations to actual throughput data from Live Starlink network, gathered over a year. Our real-world dataset, as in Fig.\ref{fig:weather}, indicates a decrease in actual downlink throughput to the user terminal on rainy days compared to clear-sky days. Simulating these conditions with \simulator yields results remarkably close to reality: within 2\% for clear-sky and within 16\% for rainy conditions. The additional discrepancy for rainy weather arises because \simulator uses ITU models to estimate weather impacts, which may not capture the full extent of real-world weather effects. Despite this, \simulator still provides a reasonable approximation of how weather conditions can affect network performance, not captured by Hypatia or StarryNet.

~\vspace{-0.20in}
\subsection{CPU and Memory Footprint}
\begin{figure}[t] %h
  \centering
  \subfigure[]
  {
    \includegraphics[width=0.465\linewidth]{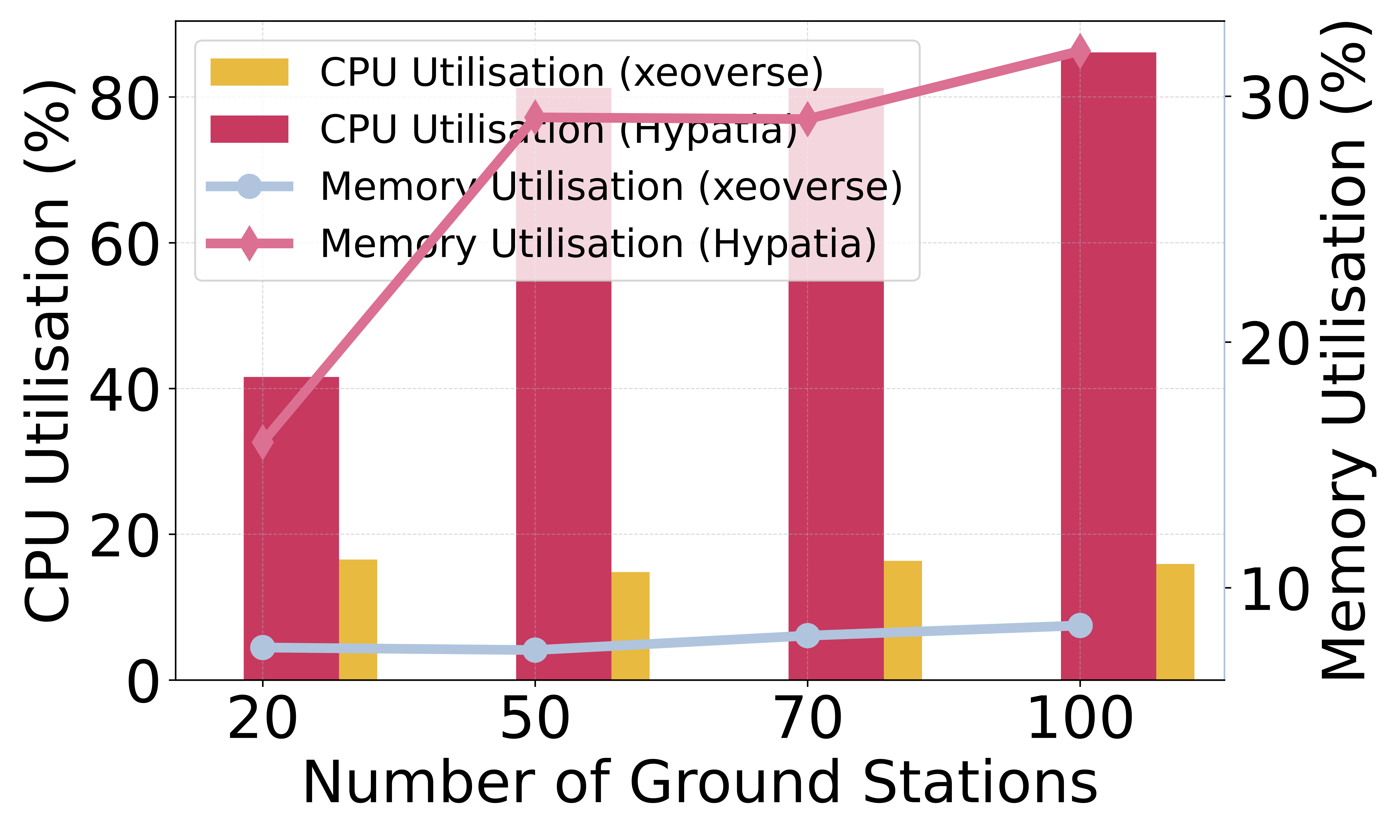}
  } 
  \subfigure[]
  {
    \includegraphics[width=0.465\linewidth]{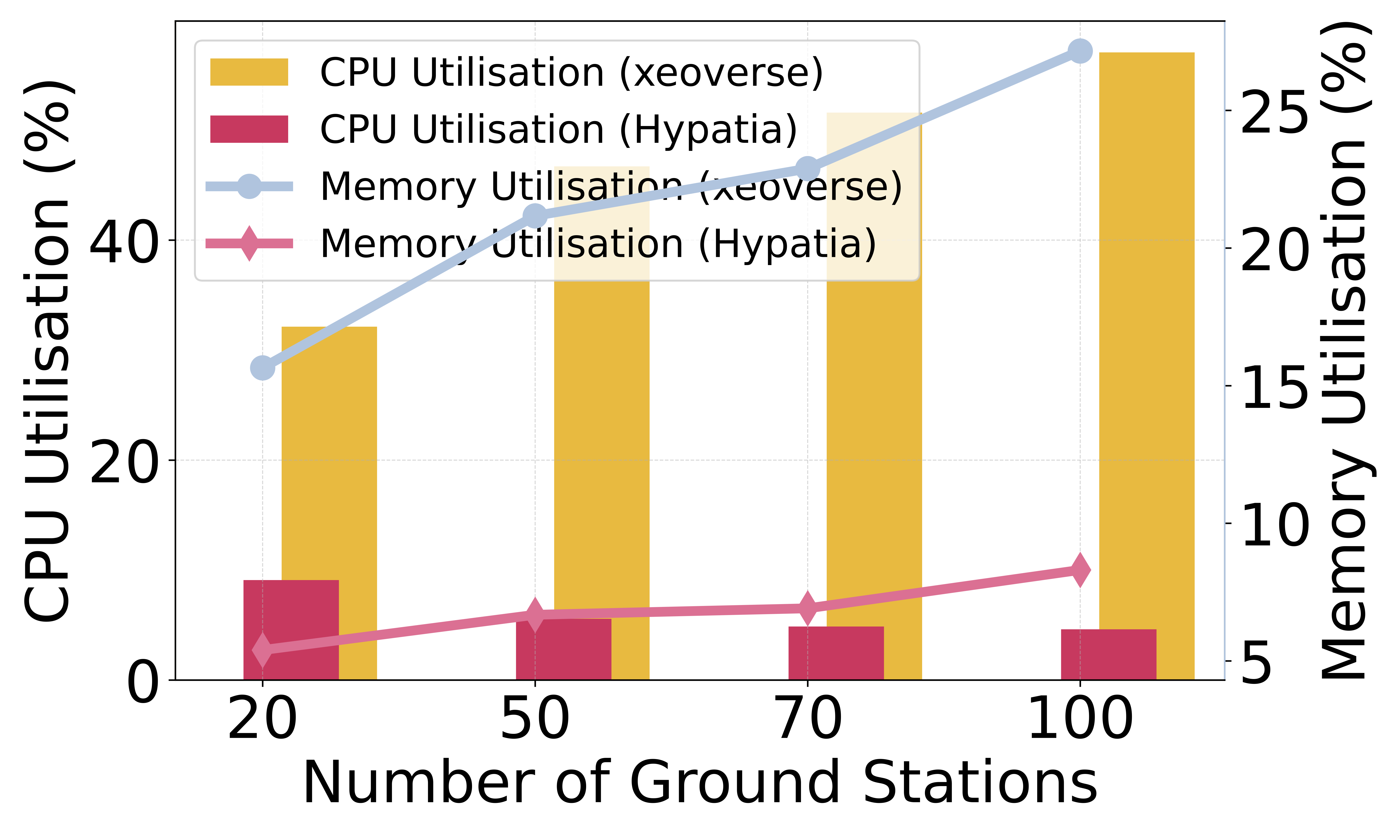}
  }
\vspace{-.1in}  
\caption{\textmd{(a) \simulator shows a lower CPU and Memory utilization compared to Hypatia in the simulation phase. (b) In pre-computation phase (runs only once per simulated scenarios) \simulator shows higher CPU and Memory utilization compared to Hypatia}}
\vspace{-.24in}
  \label{fig:footprint}
\end{figure} %h

In the third set of experiments, we simulate the full constellation of $1584$ satellites, focusing on the CPU and memory usage of both Hypatia and \simulator during the pre-computation and simulation update phases (we omit StarryNet for the same reason as before -- it is unable to scale beyond $\approx$ 1012 satellites). Fig.\ref{fig:footprint}(a) illustrates the CPU and memory consumption during the simulation update phase, while Fig.\ref{fig:footprint}(b) highlights the CPU and memory footprint for the pre-computation phase. During the pre-computation phase, and with $100$ ground stations, \simulator's CPU usage spikes to $57\%$, which is higher, $12$x times, than that of Hypatia. However, this scenario reverses during the simulation update phase: \simulator's CPU usage drops to $15.9\%$, making it $5.4$x times more efficient than Hypatia in terms of CPU consumption. Similarly, when it comes to memory usage, \simulator proves to be more efficient during the simulation update phase, using only $8.5\%$ of memory compared to $31.8\%$ for Hypatia. The results highlight that \simulator has a low resource footprint during the simulation updates phase, which is advantageous as this phase is often repeated (re-run) multiple times for various purposes, including ensuring statistical validity or comparing different algorithms, such as TCP congestion control variants, within a single scenario. The higher CPU usage observed during the pre-computation phase of \simulator is less critical, given that this phase is executed only once.

~\vspace{-0.15in}
\section{Conclusion}

We introduced \simulator, a simulator for LEO satellite mega-constellation networks designed to achieve scalability, responsiveness, high fidelity, and a low footprint. We built \simulator based on three key observations: (i) LEO satellites are predictable, allowing us to pre-compute changes in topology and network routes; (ii) at any given time step, there are only a small number of ISL changes in a mega-constellation, therefore, updating all the ISL links is unnecessary; (iii) focusing on links crucial to the simulation scenarios can significantly reduce computational demand. \simulator outperforms state-of-the-art simulators Hypatia and StarryNet.

~\vspace{-0.2in}

\bibliographystyle{IEEEtran}  
\bibliography{references}
\end{document}